\documentclass[%
 reprint,superscriptaddress,groupedaddress, longbibliography,
 amsmath,amssymb,prb,
 aps]{revtex4-2}    

\usepackage{amsmath,amssymb}
\usepackage{graphicx}
\usepackage{xcolor}
\usepackage{dcolumn}
\usepackage{bm}
\usepackage{bbold}
\usepackage{hyperref}
\usepackage{siunitx}
\usepackage{bibunits}
\defaultbibliographystyle{apsrev4-2}
\defaultbibliography{bibfile}
\newcommand{\TUdo}{Condensed Matter Theory, Department of Physics, TU Dortmund, 44221 Dortmund, Germany}

\begin{document}

\begin{bibunit}[apsrev4-2]
\preprint{arxiv}
\title{Numerically exact simulation of open quantum networks with strong system-bath couplings using comb tensor network path integrals}
\author{Quentin W. Richter}
\email{quentin.richter@tu-dortmund.de}
\affiliation{\TUdo}    
\author{Moritz Cygorek}
\email{moritz.cygorek@tu-dortmund.de}
\affiliation{\TUdo}

\begin{abstract}
Non-Markovian open quantum networks are among the most challenging problems 
in quantum theory, as they suffer from both, 
the exponentially scaling many-body state space 
as well as long memory times when the sites are strongly coupled to 
structured environments.
Recent efforts to address them by combining tensor networks with 
real-time path integrals, such as TEMPO and process tensors, 
were however limited to systems with rather weak system-environment couplings 
because strong couplings lead to large bond dimensions.
Here, we devise a tensor network scheme (CTEMPO) resulting in particularly small 
bond dimensions for single-site problems. We then extend this to a practical scheme to
simulate non-Markovian quantum networks by time-evolving a corresponding comb tensor network. 
Benchmarking the single-site algorithm on the spin-boson model we find orders-of-magnitude 
improvement over established methods.
The multi-site algorithm is shown to solve the notoriously difficult problem of
the 7-site Fenna-Matthews-Olson photosynthetic complex, whose spectral density contains 62 peaks.
We then apply our approach to explore the scaling behavior of the 
superradiance of up to $N=20$ quantum dots individually coupled to super-Ohmic phonon baths.
Our approach provides a turnkey solution to non-Markovian quantum networks
and is compatible with more general tensor network topologies. 
\end{abstract}
\maketitle
\textit{Introduction}---%
Condensed matter quantum systems including superconducting 
qubits~\cite{PT_White}, 
semiconductor quantum nanostructures~\cite{Cosacchi_QRT}, 
and biological energy transfer complexes~\cite{Lorenzoni2025}
are often strongly coupled to their environment. 
This manifests itself in non-Markovian memory effects, which probe the 
quantum dynamical evolution of the environment in addition to that of
the system. Thus,
advanced methods beyond Markovian master equations~\cite{deVega} are needed
to correctly assess the impact of the environment on the system and
on its performance as quantum device~\cite{bracht2023temperature}.
For strong couplings to strongly structured environments, even
the simulation of a single few-level system is challenging. 
The complexity increases drastically, when a number of such 
single condensed matter units are combined into, e.g., into
biological~\cite{Lorenzoni2025} or bio-inspired~\cite{guide_slide} 
energy transfer networks.

\begin{figure*}
\includegraphics[width=0.99\textwidth]{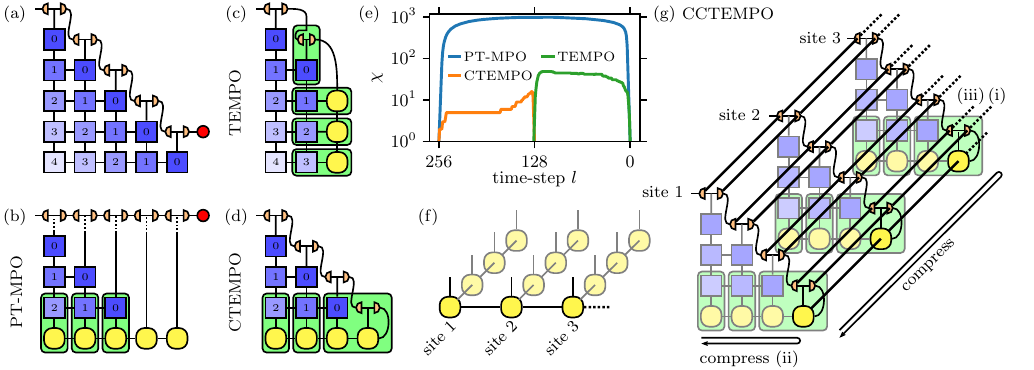}
\caption{\label{fig:sketch}(a) Tensor network representation of the path summation in Eq.~\eqref{eq:pathsum}. 
The initial state (red circle) 
is propagated by contracting with system propagators (orange semicircles)
and influence tensors (blue boxes).
Different methods (b-d) correspond to different contraction orders. The current MPO (yellow boxes) is updated by including, in the next step, the terms contained in the green boxes.
In TEMPO (c) the tensor network is contracted column by column. 
PT-MPOs (b) are obtained by contracting only the influence tensors. This is done row by row to utilize
the causal structure of the tensor network~\cite{JP}. The system propagators are later contracted with the final PT-MPO.
Our CTEMPO scheme (d) also utilizes the causal structure by contracting row by row but also contracts the system
propagators into the network at an early stage to shorten the MPO and to make it more compressible. 
Panel (e) shows the typical profile of inner bond dimensions for the three methods simulating an Ohmic spin boson model for $\alpha=1$ 
(see Fig.~\ref{fig:spinboson}).
The compression threshold was chosen so that the numerical error is $\sim10^{-3}$.
(f) A multisite algorithm is constructed by elevating the MPO of a system density matrix to the backbone of a comb
whose teeth are CTEMPO MPOs [yellow boxes in (d)].
(g) CCTEMPO algorithm: The comb tensor network in (f) is updated by alternatingly propagating the backbone by contracting with 
system propagator MPOs [step (i) and (iii)] and the teeth [step (ii)] by CTEMPO updates as in panel (d).
}
\end{figure*}

Here, we demonstrate a practical numerically exact approach to 
simulate non-Markovian quantum networks with 
strong couplings to their environment
based on real-time path integrals and tensor networks.
Applying matrix product operator (MPO) representations
to path integrals was pioneered by Strathearn \textit{et al.}~\cite{TEMPO} 
with the TEMPO method, which reduces the exponential scaling of QUAPI~\cite{QUAPI}
with the memory time to polynomial.
J{\o}rgensen and Pollock~\cite{JP} suggested to contract the same tensor network 
differently to obtain an MPO representation of the Feynman-Vernon influence 
functional~\cite{FeynmanVernon}, which is then called the
process tensor (PT-MPO) [see Fig.~\ref{fig:sketch}(a-c)]. 

PT-MPOs are independent of the system propagator and can be precomputed 
and reused to efficiently explore system parameters and driving 
protocols~\cite{Fux_optimal_control}, and to obtain integrals over 
multi-time correlation functions~\cite{bracht2023temperature}. Moreover, 
multiple independent environments attached to the system can be treated 
numerically exactly by combining PT-MPOs calculated independently for each 
environment~\cite{Wiercinski1, ACE, Gribben2022}.
Naturally, this suggests extensions to non-Markovian quantum networks
by attaching PT-MPOs to the MPO representation of the many-body density 
matrix~\cite{Fux_spinchain,Bose2022}.
In practice, however, this approach is limited because
PT-MPOs for realistic environments can have sizable bond dimensions.

The inner bonds of PT-MPO represent the most relevant subspace of the
environment Liouville space for the future system evolution
if neither system Hamiltonian nor initial state are fixed~\cite{inner_bonds}. 
Consequently, large bond dimensions are expected
if a large environment Liouville subspace is explored, 
such as at high temperatures~\cite{Strunz_PT_stochastic} 
or in the ultra strong coupling regime of the (sub-)Ohmic spin-boson model~\cite{Chin_subohmic}. 
An example for which obtaining the PT-MPO is infeasible~\cite{ACE_code} 
is the Fenna-Matthews-Olson (FMO) complex with a spectral density comprising 62 peaks~\cite{Raetsep}. 
Several algorithms for the computation of PT-MPOs have been 
developed to deal with large memory and propagation 
times~\cite{DnC,LinkStrunz}, larger system sizes~\cite{Cochin2026}, and
environments beyond the spin-boson model~\cite{ACE,Abanin} (see reviews in
Refs.~\cite{ReviewKeeling,ReviewLacroix}). However, the inner bond dimension of the final
PT-MPO is largely independent of the algorithm used
to obtain it~\cite{inner_bonds}, and so novel
PT-MPO algorithms are not expected to solve the fundamental problem of large bond dimensions.

To achieve much smaller bond dimensions, it is suggestive to contract
information about the system propagation into the tensor network,
as it is in TEMPO~\cite{TEMPO}. 
However, J{\o}rgensen and Pollock also pointed out that the 
contraction order of TEMPO does not exploit the causal order and thus produces
unnecessarily large intermediate bond dimensions~\cite{JP}.
Here, we first combine both insights  
to devise a causal TEMPO algorithm (CTEMPO) retaining very small bond 
dimensions.

\textit{Single-site method CTEMPO}---%
We consider the spin-boson model $H=H_S+H_E$ with a general system part
$H_S$ and environment part
\begin{align}
\label{eq:spinboson}
H_E =& \hbar \sum_k\bigg(\omega_{k} b^\dagger_{k}b_{k}
+\big(g_{k}b^\dagger_{k}+g^*_{k}b_{k} \big)\hat{O}\bigg),
\end{align}
where $b^\dagger_{k}$ and $b_{k}$ are creation and
annihilation operators of the $k$-th boson mode with frequency
$\omega_{k}$ coupled to the system site with coupling constant 
$g_{k}$. $\hat{O}$ is a Hermitian coupling operator.
The environment is characterized by the spectral density 
$J(\omega)=\sum_{k}|g_{k}|^2\delta(\omega-\omega_{k})$.

Real-time path integrals utilize the Gaussian nature of the harmonic oscillator 
environment, i.e. only second-order correlations contribute. 
Using the short-hand notation
$\bar{\rho}_{\alpha_l}=\bar{\rho}_{\nu_l,\mu_l}(t_l)$ for the reduced system
density matrix elements at time step $t_l=l\Delta t$ with double index 
$\alpha_l=(\nu_l,\mu_l)$, the time evolution is given by~\cite{JP}
\begin{align}
\label{eq:pathsum}
\bar{\rho}_{\alpha_n}=\sum_{\{\alpha_{l}\}_{l=0}^{n-1}} 
\bigg(\prod_{j=1}^n \prod_{i=j}^{n} b_{i-j}^{\alpha_i,\alpha_j} \bigg)
\bigg(\prod_{l=1}^n\mathcal{M}_{\alpha_l,\alpha_{l-1}}\bigg)
\bar{\rho}_{\alpha_0},
\end{align}

where $\mathcal{M}=e^{\mathcal{L}_S\Delta t}$ is the system propagator with 
system Liouvillian, e.g., $\mathcal{L}_S=-\frac{i}{\hbar}[H_S, .]$,
and 
$b_{i-j}^{\alpha_i,\alpha_j}$ describes how the system state at past time
$t_j$ affects the system at future time $t_i$ via the environmental memory
(see the Appendix~\ref{app:b}). The double product over these terms yields
the Feynman-Vernon influence functional~\cite{FeynmanVernon}.
$\bar{\rho}_{\alpha_0}$ is the system initial state and the environment is
initially thermal at temperature $T$.

The path summation in Eq.~\eqref{eq:pathsum} can be represented as a 
tensor network with nodes $b_{i-j}^{\alpha_i,\alpha_j}$ and 
$\mathcal{M}_{\alpha_l,\alpha_{l-1}}$ as shown in 
Fig.~\ref{fig:sketch}(a). 
Horizontal legs correspond to past time index $\alpha_j$ while
vertical legs represent the future index $\alpha_i$ on node 
$b_{i-j}^{\alpha_i,\alpha_j}$.
Legs between two nodes can be contracted by summation over common indices. 
To obtain the final density matrix $\bar{\rho}_{\alpha_n}$, the full tensor network has to be contracted.

All three approaches
operate on one-dimensional MPOs that are contracted with another row or column 
of the remaining tensor network using the zipup scheme~\cite{zipup}. 
To keep the inner bonds tractable, every contraction is followed by MPO
compression, which is realized by singular value decompositions (SVDs)
and truncation of small singular values 
$\sigma_i<\epsilon\sigma_0$, where $\sigma_0$ is the largest
singular value and $\epsilon$ is the compression threshold that determines
accuracy and computation time.

The three methods differ by the contraction order as shown in Fig.~\ref{fig:sketch}(b-d).
In TEMPO~\cite{TEMPO}, the MPO contains the history of the system state.
The indices at the interface between the MPO and the 
next column are past time indices.
The PT-MPO represents the full influence functional and is
contracted leaving the upper bonds dangling, before contracting
with the system propagator in a separate step. 
According to Ref.~\cite{JP} contracting the PT-MPO row-by-row instead of
column-by-column exploits the causal structure, which leads to smaller 
bond dimensions in intermediate steps.
In CTEMPO, we also contract row-by-row but, in the $l$-th step, 
we also contract with the $l$-th system propagator to eliminate the corresponding leg.
(see the Appendix~\ref{app:CTEMPO} for the explicit algorithm).
This shortens the MPO in each time step and moves the edge, where compression
is most effective, close to the short-time memory nodes, which contribute 
the most to large bond dimensions.

\begin{figure}
\includegraphics[width=\linewidth]{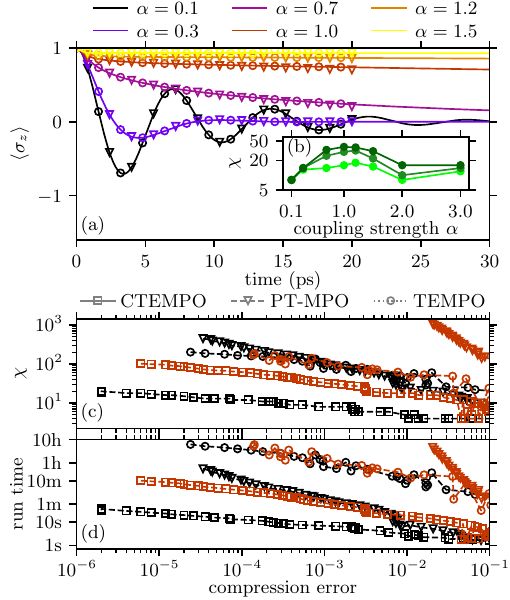}
\caption{\label{fig:spinboson}(a) Dynamics of the spin-boson model solved using
TEMPO, PT-MPO, and CTEMPO schemes.
The inset shows the maximal inner bond dimension in CTEMPO simulation as a function of the coupling strength $\alpha$ 
for final propagation times $\Omega t_f = 10.24,\,20.48,\,40.96$ (from light to dark green).
(c) Bond dimensions and (d) computation times on an Intel Xeon w3-2435 processor, respectively, as a function of the 
compression error for $\Omega t_f=20.48$ and for two values of the coupling
strength $\alpha=0.1$ (black) and $\alpha=1$ (brown).
}
\end{figure}

We compare these methods on the Ohmic spin-boson model
with $H_S=(\hbar\Omega/2)\sigma_x$ and 
$J(\omega)=2\alpha \omega e^{-\omega/\omega_c}$ at $T=0$~K. 
The time evolution of $\langle \sigma_z\rangle$ after preparing the state
$|\uparrow\rangle$ is shown in Fig.~\ref{fig:spinboson}(a) 
for various values of $\alpha$. 
All three methods agree and reproduce the established 
findings~\cite{Leggett,Bulla2003} of a transition 
between underdamped and overdamped dynamics at $\alpha_c\approx 0.5$ 
and a transition to a localized phase with nonzero
stationary $\langle\sigma_z\rangle$ at a critical value $\alpha_c\gtrsim 1$.

We vary the compression threshold at fixed time step $\Omega\Delta t=0.04$ and 
compare the computation time and the maximal bond dimension $\chi$ after MPO compression occurring in the different algorithms as a function of the compression error in Fig.~\ref{fig:spinboson}(d) and (c), respectively.
We find that PT-MPOs are faster than TEMPO for weak coupling $\alpha=0.1$ (due to the causal order), whereas for larger
coupling $\alpha=1$ the larger bond dimensions make PT-MPOs significantly 
slower. In both cases, CTEMPO is one to two orders of magnitude faster than both, TEMPO and the PT-MPO approach. 
This speed-up stems from the particularly small bond dimensions $\chi$ of CTEMPO [see also the bond dimension profile in Fig.~\ref{fig:sketch}(e)] and the scaling of the computation time of SVDs as $\mathcal{O}(\chi^3)$. 
While PT-MPOs acquire bond dimensions of hundreds to thousands, CTEMPO yields accurate results with single-digit or low two-digit values of $\chi$.
As shown in the inset in Fig.~\ref{fig:spinboson}, the bond dimensions even remain manageable for long propagation times 
and across the localization phase transition at $\alpha\sim 1$. We interpret the bond dimension maximum at this point as indication that temporal correlations, much like spatial correlations~\cite{Vidal_critical}, grow close to phase transitions.

\textit{CCTEMPO method for quantum networks}---%
Numerically exact methods for multi-site quantum networks have been proposed based on PT-MPOs~\cite{Fux_spinchain}. However, as discussed above, PT-MPOs may have large bond dimensions. A multi-site version of CTEMPO is thus a more likely candidate for a practicable numerically exact solver for non-Markovian quantum networks. 
Our starting point is an MPO representation of the density matrix of a many-body quantum system without non-Markovian environments, which can be evolved in time by MPO-MPO multiplications with an MPO representation of
the system propagator (see Supplementary Material~\ref{SM:MPOMPO}).
For a system whose sites are coupled to non-Markovian environments, the MPO matrices of
the many-body density matrix are replaced by corresponding CTEMPO MPOs. 
The result is the comb tensor network depicted in Fig.~\ref{fig:sketch}(f).

In our comb-CTEMPO (CCTEMPO) algorithm [Fig.~\ref{fig:sketch}(g)], the comb tensor network is updated 
by propagating the system over half a time step 
via MPO-MPO multiplication on the backbone before and after 
CTEMPO updates are applied to each tooth. 
The computational bottleneck of CCTEMPO are the updates of the backbone.
Compared to the propagation of a Markovian open quantum system,
the extra legs connecting the backbone to the teeth increase the 
computation time for MPO-MPO updates linearly by $\chi$. Since $\chi$ remains
small in CTEMPO, our approach remains practicable even for strong
couplings to structured environment.

\begin{figure}
\includegraphics{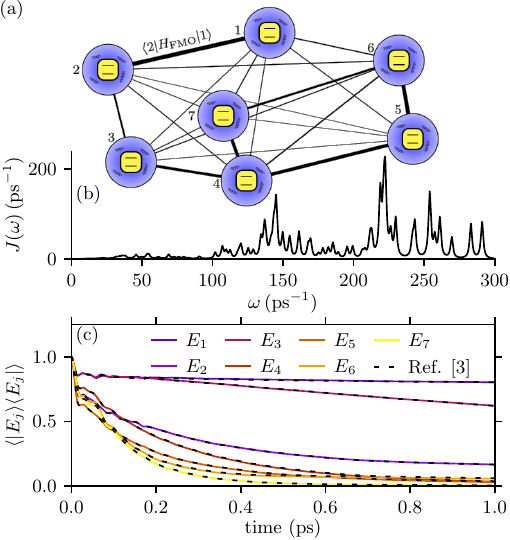}
\caption{\label{fig:FMO}(a) 7-site Fenna-Matthews-Olson complex as a non-Markovian quantum network, where
each site is coupled to a strongly structured phonon bath at $T=77$ K with spectral density in (b). 
(c) Dynamics of initially populated energy eigenstates for $\Delta t=1$ fs and $\epsilon=10^{-6}$.
Dashed lines show reference results obtained using DAMPF~\cite{Lorenzoni2025}. 
}
\end{figure}
We demonstrate this on the example of the Fenna-Matthews-Olson (FMO) 
photosynthetic complex with the 7-site Hamiltonian of Ref.~\cite{AdolphsRenger} 
[Fig.~\ref{fig:FMO}(a)]
and a bath with the 62-peak spectral density of Ref.~\cite{Raetsep} 
[Fig.~\ref{fig:FMO}(b)] attached at each site (see Supplementary Material~\ref{SM:FMO}).

This model is particularly challenging because of
the similar energy scales of system and environment interactions,
the long memory times due to the sharp peaks in the spectral densities,
and the large number of distinct peaks, which imply a large number of 
dynamical degrees of freedom.
Finally, high (biological) temperatures pose challenges for 
wave-function-based methods and result in strong effective system-environment
couplings.
Such problems have only recently been successfully addressed
using methods based on path summation truncating temporal 
correlations~\cite{Kundu2022} or the DAMPF~\cite{Lorenzoni2025,
DAMPF} method, where a set of damped harmonic oscillators is fitted to
the bath correlation function, which is then evolved using MPOs.
Notably, these approaches consider only the 
single excitation manifold as a single multi-level system, whereas
we deliberately simulate a non-Markovian quantum network, 
where each site is a two-level system.

\begin{figure}
\includegraphics[width=\linewidth]{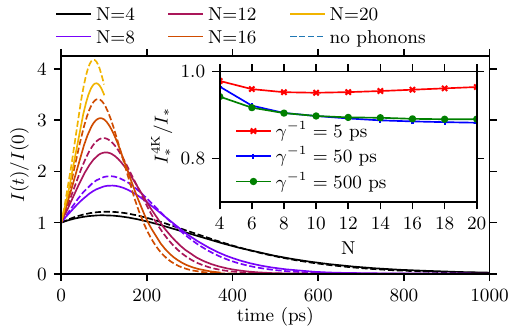}
\caption{\label{fig:superrad}Superradiant emission of up to $N=20$ semiconductor 
quantum dots coupled to individual super-Ohmic phonon baths. The radiative lifetime is $\gamma^{-1}=500$ ps.
The inset shows the ratio between
the maximal peak intensities for simulations with and without accounting for phonons.
}
\end{figure}
Despite this much larger state space, CCTEMPO can be used directly 
and without any fine tuning to calculate the decay of initially 
prepared eigenstates of the FMO Hamiltonian [see Fig.~\ref{fig:FMO}(c)].
Simulations typically require 3-17 hours 
to converge within 
$1\%$ to DAMPF results~\cite{Lorenzoni2025}. 

To demonstrate the versatility of CCTEMPO, we 
now explore how local super-Ohmic phonon baths affect the superradiance of
quantum emitters. This model is motivated by recent progress in fabricating
semiconductor quantum dots (QDs) with collective coupling to a common 
waveguide~\cite{superrad_Kim,superrad_Hallett}. 
Recent experiments on cooperative emission of up to 5 QDs~\cite{Shaji2026} demonstrate the need for corresponding theoretical methods.
Numerically exact simulations
of two QDs~\cite{Wiercinski1,Wiercinski2} reveal that the effects of
super-Ohmic phonon baths is qualitatively different from Ohmic baths or 
local Lindbladian dephasing, as coherences initially drop on the timescale
of the bath memory time and then only decay very slowly via modified collective decay rates.
This example is quite different from the FMO complex due to the different
bath spectral density~\cite{Krummheuer2005} as well as the system propagation,
which is governed by a single Lindblad term with a collective collapse operator 
$L=\sum_{i=1}^N\sigma^-_i$
(see Supplementary Material~\ref{SM:superrad}).
We also use memory truncation to achieve linear scaling in the propagation
time $t_f$. Note that the reduced system state is 
highly mixed and composed of states with strong correlations across the
whole network, which makes even a Markovian treatment numerically challenging.  

Nevertheless, CCTEMPO delivers numerically complete simulations of up $N=20$ QDs,
which is large enough to infer insights about the large-$N$ scaling behavior,
as shown in Fig.~\ref{fig:superrad}. 
Models using local Lindblad dephasing~\cite{Bassler2026} predict that 
collective decay dominates over local dephasing and the ratio between the 
peak burst intensities in simulations with and without dephasing
approaches 1 for $N\to \infty$. By contrast, our simulations indicate that local super-Ohmic
phonon coupling leads to a nearly constant ratio $I^{4K}_*/I_*$ for 
large $N$ for realistic parameters with exciton lifetimes $\gamma^{-1}=50-500$ ps.
However, care must be taken when extrapolating this behavior to 
$N\to \infty$, because the time $\tau_d$ at which the peak intensity is 
maximal decreases with increasing $N$ as $\tau_d\sim \ln N/(N\gamma)$. 
For an artificially small lifetime $\gamma^{-1}=5$ ps, 
the delay time $\tau_d$ becomes smaller than the environment memory time
of about $3$~ps, and so the emission is faster than the 
phonon-induced dephasing.
Hence, our numerically complete analysis lets us discover a novel intermediate scaling 
regime of the superradiant burst intensity in the presence of non-Markovian
environments.

\textit{Conclusion}---%
To summarize, we demonstrate a path integral tensor network method with
particularly small bond dimensions that not only leads to orders-of-magnitude improvement in single-site non-Markovian quantum system simulation but also yields a practicable numerically exact approach for non-Markovian quantum networks.
Our examples of
the FMO complex and quantum dot superradiance demonstrate its 
practicality even in the most adversarial cases of strong coupling 
to strongly structured environments and for networks with all-to-all
connectivity. 
The only constraint is that local environments are spin-boson-like,
but generalizations to other Gaussian 
environments~\cite{Guo_generalGaussian,Link_generalGaussian} are foreseeable.
We derived the comb topology starting from a MPO description of the quantum network, which is ideal for quasi-onedimensional systems.
Recently, belief propagation~\cite{BeliefPropagation_SciPost,BeliefPropagation_Chan,BeliefPropagation_Guo} enabled the contraction of higher dimensional quantum systems.
Such system can similarly be elevated to include non-Markovian
environments by attaching CTEMPO teeth at each site. This does 
not introduce loops in the network and increases the 
cost only linearly in the (small) CTEMPO bond dimension $\chi$.
For systems well represented by MPOs, CCTEMPO provides a turnkey
solution 
that requires no fine tuning other then choosing suitable convergences parameters $\Delta t$ and $\epsilon$.

\textit{Acknowledgements}---%
The authors thank Nicola Lorenzoni for providing reference data on the FMO complex. M.C. is supported by the Return Program of the State of North Rhine-Westphalia.

\textit{Data availability}---%
The CTEMPO and CCTEMPO methods and parameters for our examples are available within the ACE code~\cite{ACE_github}.

\appendix

\section{\label{app:b}Influence functional terms}
To obtain the terms $b_{i-j}^{\alpha_i,\alpha_j}$ in the influence functional,
we first unfold the double indices $\alpha_l=(s_l,r_l)$ into forward and 
backward Hilbert space indices $s_l$ and $r_l$, respectively. Furthermore,
we work in the eigenbasis of the coupling operator $\hat{O}$ in 
Eq.~\eqref{eq:spinboson}, whose eigenvalues are denoted $\lambda_{s_l}$ and
$\lambda_{r_l}$, respectively.

Using the bath correlation function
\begin{align}
C(\tau)=\int\limits_0^\infty d\omega\, J(\omega)\bigg[
\coth\bigg(\frac{\hbar\omega}{2k_BT}\bigg)\cos(\omega \tau) 
-i\sin(\omega\tau)\bigg]
\end{align}
and its time integrals
\begin{align}
\eta_{l}=\begin{cases}
\int\limits_{0}^{\Delta t} dt'\int \limits_{0}^{t'}dt''\,
C(t'-t''),&\text{ for } l=0,\\
\int\limits_{l\Delta t}^{(l+1)\Delta t} dt'\int \limits_{0}^{\Delta t}dt''\,
C(t'-t''),&\text{ for } l\neq 0,\\
\end{cases}
\end{align}
the influence tensors in the tensor network are given by
\begin{align}
b_{i-j}^{\alpha_i,\alpha_j}=e^{-(\lambda_{s_i}-\lambda_{r_i})
(\eta_{i-j}\lambda_{s_j}-\eta^*_{i-j}\lambda_{r_j})}.
\end{align}

\section{\label{app:CTEMPO}CTEMPO algorithm}
The CTEMPO algorithm operates on an MPO of length $n_c$, which is the
memory cutoff time, or $n_c=n$ the total number of time steps
if no memory truncation is used.
The MPO matrices are initialized by the system initial state in the first
node
$\mathcal{Q}^{\alpha_0}_{d_0,0} = \delta_{d_0,0}\rho_{\alpha_0}$
and with 
$\mathcal{Q}^{i(\alpha_l)}_{d_l, d_{l-1}}=\delta_{d_l,0}\delta_{d_{l-1},0}$ 
in the remaining MPO sites. The notation $i(\alpha_l)$ is used to accommodate
the reduction of the bond dimension by exploiting degeneracies in the spectrum 
of the coupling operator~\cite{cavity_feeding}.
$i(\alpha_l)$ maps different indices $\alpha_l$ belonging to 
same degenerate eigenvalue to the same outer bond index. If no such 
degeneracy is present, one can set $i(\alpha_l)=\alpha_l$.

In every iteration steps, the following updates are performed:
For the first site
\begin{align}
\mathcal{Q}^{\alpha'_0}_{(d_1,\beta),0}=
\sum\limits_{\alpha_0,\alpha_1,d_1} 
\mathcal{M}^{1/2}_{\alpha'_0,\alpha_1}
b_0^{\beta,i(\alpha_1)} \mathcal{Q}^{i(\alpha_1)}_{d_1,d_0} 
\mathcal{M}^{1/2}_{\alpha_1,\alpha_0} \mathcal{Q}^{\alpha_0}_{d_0,0},
\end{align}
where $\mathcal{M}^{1/2}=e^{\mathcal{L}_S\Delta t/2}$ is the system 
propagator over half a time step (to achieve a symmetric 
Trotter splitting). 

All but the first and the last matrix are updated as
\begin{align}
\mathcal{Q}^{i(\alpha'_l)}_{(d_{l+1},\beta),(d_{l},\beta)}=
\sum_{i(\alpha_{l+1})}\delta_{i(\alpha'_l), i(\alpha_{l+1})}
b_l^{\beta,i(\alpha'_l)} \mathcal{Q}^{i(\alpha_{l+1})}_{d_{l+1},d_l}.
\end{align}

If no memory truncation is employed $n_c=n$, the last MPO matrix is 
dropped---shortening the chain---and the second bond index of the 
penultimate matrix is traced out 
$\sum_{d_{l+1},\beta}q_{d_{l+1},\beta} 
\mathcal{Q}^{i(\alpha'_l)}_{(d_{l+1},\beta),(d_{l},\beta)}$ 
using the closure  $q_{d_{l+1},\beta}$ (see below).
If the memory is truncated, the MPO length is not shortened in the first 
$n-n_c$ iterations, and instead the last MPO matrix is set to
$\mathcal{Q}^{i(\alpha_{n_c})}_{0,\beta}=b_{n_c}^{\beta,i(\alpha_{n_c})}$.

The MPO is compressed already during the update, following the 
zipup scheme~\cite{zipup}. 
The compression is realized by a sweep with truncated singular value 
decompositions with relative compression threshold $\epsilon$. 
The zipup sweep is followed by a backward sweep 
with the same compression threshold.

The reduced system density matrix at a given iteration step is obtained by
$\rho_{\alpha'_0}=\sum_{d_0} q_{d_0} \mathcal{Q}^{\alpha'_0}_{d_0,0}$ using 
the closure $q_{d_l}$, which is equivalent to taking the trace over the 
environment degrees of freedom~\cite{ACE}. 
Here, we obtain $q_{d_l}$ by initializing 
it as $q_{d_l}=\delta_{d_l,0}$ and updating it whenever the MPO matrices
$\mathcal{Q}$ are updated and compressed, as described in 
Ref.~\cite{inner_bonds} for PT-MPOs.

%

\end{bibunit}

\clearpage 
\newpage

\makeatletter
\def\appendixname{} 
\makeatother

\setcounter{page}{1}       
\setcounter{section}{0}       
\setcounter{equation}{0}   
\setcounter{figure}{0}     
\setcounter{table}{0}      
\renewcommand{\thesection}{S\arabic{section}}   
\renewcommand{\theequation}{S\arabic{equation}}
\renewcommand{\thefigure}{S\arabic{figure}}
\renewcommand{\thetable}{S\arabic{table}}

\begin{bibunit}[apsrev4-2]
  
\renewcommand{\bibnumfmt}[1]{[S#1]} 
\renewcommand{\citenumfont}[1]{S#1} 
\onecolumngrid 
\begin{center}
    \textbf{\large Supplementary Material: Numerically exact simulation of open quantum networks with strong system-bath couplings using comb tensor network path integrals} \\
    \vspace{0.3cm}
    Quentin W. Richter and Moritz Cygorek\\
    \textit{\TUdo}
\end{center}
\vspace{0.6cm}

\section{\label{SM:spinboson}Computation time and compression errors for the 
single-site spin-boson model}
\begin{figure*}[h]
\includegraphics[width=\linewidth]{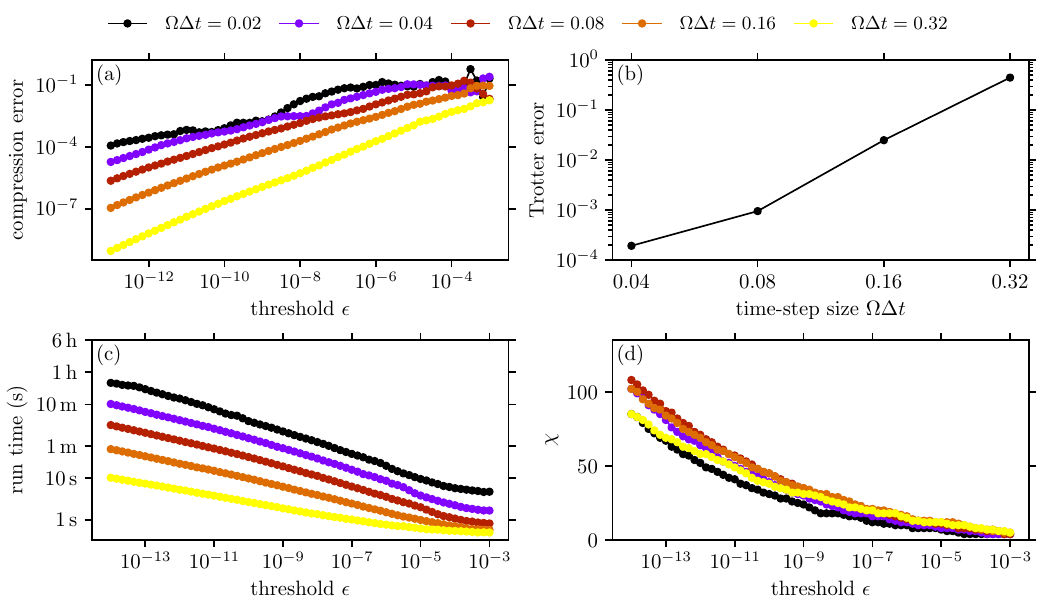}
\caption{\label{fig:CTEMPO_convergence}
Convergence study of CTEMPO for the Ohmic spin-boson model, discussed in the main text, with $\alpha=1$ and $\Omega t_e=20.48$ .
The panels show the compression error (a), run time (c), and maximal inner bond dimension (d) as a function of the compression threshold for different time discretizations. (b) shows the error due to time discretization.
}
\end{figure*}
Throughout this letter, we define the compression error as the maximal
absolute deviation of $\langle \sigma_z \rangle$ at any time step compared to a well converged reference simulation, i.e.,
\begin{equation}
E(\Omega\Delta t,\,\epsilon) = \underset{l\in[0,n]}{\mathrm{arg\,max}} \,|\langle \sigma_z \rangle(t_l;\;\epsilon,\,\Omega\Delta t) - \langle \sigma_z \rangle(t_l;\;\epsilon_{\mathrm{ref}},\,\Omega\Delta t)|.    
\label{eq:num_err}
\end{equation}
Unless stated otherwise, all simulations are done at $\Omega\Delta t=0.04$,
whose Trotter error can be estimated to be (vide infra) $\sim 10^{-4}$.
Numerical errors shown in Fig.~\ref{fig:spinboson} (c) and (d) in the main text
are computed w.r.t. to CTEMPO simulations at $\epsilon_{\mathrm{ref}}=10^{-15}$.
The compression errors of TEMPO and the PT-MPO method can be computed with respect to the same reference simulation since all three methods contract the same tensor network, and therefore, have the same Trotter error.
The thresholds and errors to obtain the profiles shown in Fig.~\ref{fig:sketch} (e) are
$E(0.04,\,\epsilon_{\mathrm{CTEMPO}}=10^{-7})=0.00092$, $E(0.04,\,\epsilon_{\mathrm{TEMPO}}=1.02\cdot10^{-8})=0.00143$, and
$E(0.04,\,\epsilon_{\mathrm{PT-MPO}}=10^{-11})=0.00152$.
Errors shown in Fig.~\ref{fig:spinboson} (b) are obtained by varying $\epsilon$ until the compression error is $\sim 10^{-3}$.

The simulation cost is either measured in run time (wall clock time) of the respective simulation, or the maximal inner bond dimension $\chi$ occurring after compression in any step of the algorithm.

The numerical error of the reference simulation is estimated as follows.
First, the compression error is estimated by comparing simulations with the same time discretization, i.e., same Trotter error [see Fig.\ref{fig:CTEMPO_convergence} (a)].
It turns out that the compression error is smaller for larger time steps at 
the same $\epsilon$, which is well known in the context of PT-MPO compression~\cite{ACE_code}.
More importantly, for all values of $\Omega\Delta t$ the compression error approaches $10^{-4}$ as the compression threshold 
approaches $\epsilon_{\mathrm{ref}}$.
Next, we compare simulations with different time steps that are converged w.r.t. $\epsilon$ to estimate the Trotter error [see Fig.\ref{fig:CTEMPO_convergence} (b)].
Here, we find a Trotter error of $\sim 10^{-4}$ at $\Omega\Delta t =0.04$.
The total numerical error is the sum of Trotter and compression error which is $\sim 10^{-4}$.
Note that the example of $\alpha=1$ is close to the phase transition, therefore, the error of the reference simulation is among the largest between the studied $\alpha$'s. 


\section{\label{SM:MPOMPO}Many-body system propagation in multi-site systems}
We decompose the time evolution of the comb tensor network formally 
into system and environment propagation via the symmetric Trotter decomposition
\begin{align}
e^{\mathcal{L}\Delta t} =&  e^{(\mathcal{L}_S+\mathcal{L}_E)\Delta t}
=e^{\mathcal{L}_S \frac{\Delta t}{2}}e^{\mathcal{L}_E\Delta t}
e^{\mathcal{L}_S \frac{\Delta t}{2}} +\mathcal{O}(\Delta t^3),
\end{align}
where $\mathcal{L}_S$ and $\mathcal{L}_E$ denote the system and 
environment Liouvillians. Application of the environment propagator $e^{\mathcal{L}_E\Delta t}$ is equivalent to applying a single CTEMPO
method without the multiplication with $\mathcal{M}$. 
Here, we describe how we perform the system propagation with
$\mathcal{M}^{\frac 12}=e^{\mathcal{L}_S \frac{\Delta t}{2}}$ when the
system Liouvillian contains many-body terms which couple different
sites of the quantum network.

To accommodate the most general quantum network topology, we perform 
system updates using contractions of the comb backbone with the 
system Liouvillian represented as one or multiple MPOs. 
For example, TEBD for systems with nearest-neighbors coupling can be represented as the sequential multiplication connecting pairs starting
with even and odd number sites, respectively. Moreover, if the system propagation is purely unitary, it is advisable to split up the 
Liouville space propagation into forward and backward propagation with
the respective unitary time evolution operators $e^{\mathcal{L}_S\frac{\Delta t}2}=e^{-\frac{i}{\hbar}H_S \frac{\Delta t}{2}} \otimes e^{\frac{i}{\hbar}H_S \frac{\Delta t}{2}}$. 
The numerical efficiency depends on a trade-off between the bond
dimension of the propagator MPO, which is multiplied with that of the
state before MPO compression, and the number of times a backbone-propagator multiplication has to be performed. 

Here, we consider a general system Liouvillian
\begin{align}
\mathcal{L}_S [\hat{\rho}] =& -\frac i\hbar [H_S\hat{\rho} - \hat{\rho} H_S] 
+\sum_j \gamma_j \big[ L_j \hat{\rho} L_j^\dagger 
-\frac 12(L_j^\dagger L_j\hat{\rho} + \hat{\rho} L_j^\dagger L_j)\big]
\nonumber\\
\cong& \bigg(-\frac i\hbar [H_S\otimes \mathbb{1} - \mathbb{1}\otimes H_S^T]
+\sum_j\gamma_j \big[ L_j \otimes L_j^* 
-\frac 12( L_j^\dagger L_j \otimes \mathbb{1} 
+\mathbb{1}\otimes  (L_j^\dagger L_j)^T)\big]\bigg)\rho, 
\end{align}
where, in the second line, we introduce a Liouville space
superoperator notation, in which propagation becomes a matrix-vector
multiplication with the density matrix as a vector. 
We build the propagator MPO by decomposing the system Hamiltonian into
$n$-body terms. For example, if a system Hamiltonian consists of at
most two-body terms, we can express its forward contribution to the 
Liouville superoperator by 
\begin{align}
H_S \otimes \mathbb{1} = \sum_k A_k^{(i)} B_k^{(j)},
\end{align}
where $A_k^{(i)}$ and $B_k^{(j)}$ superoperators acting only on
sites $i$ and $j$, respectively. This contribution can be written 
as an MPO with matrices
\begin{align}
\label{eq:singleterm}
W^{\alpha'_l,\alpha_l}_{0,0} = \begin{cases}
\delta_{\alpha'_l,\alpha_l}, & l\neq i, l\neq j,\\
\big(A_k^{(i)}\big)_{\alpha'_l,\alpha_l}, & l=i\neq j,\\
\big(B_k^{(j)}\big)_{\alpha'_l,\alpha_l}, & l=j\neq i,\\
\end{cases}
\end{align}
with inner bond dimension 1. Analogous expression can be used
for elementary one- or multi-site contributions.
Regarding the Lindblad terms, we note that even collective 
decay operators, such as in ideal superradiance 
$L_j=(\sigma^-_1+\sigma^-_2+\dots+\sigma^-_N)$, lead at most to 
two-body terms when products
of $L^\dagger_j$ and $L_j$ are expanded, and are thus captured by Eq.~\eqref{eq:singleterm}.

If $N_H$ MPOs from individual Hamiltonian or Lindblad terms are added,
one obtains a MPO with inner bond dimension $N_H$, which can be
compressed using trunctated SVDs. The remaining task is to
exponentiate the Liouvillian MPO to obtain the system propagator 
$e^{\mathcal{L}_S\frac{\Delta t}2}$ in MPO form.
We find it practicable to obtain this exponential by constructing a 4-th order Taylor expansion using MPO-MPO additions and
multiplications, where for the latter the zipup scheme is used and 
sweeps with SVDs are employed to keep the bond dimension minimal.

For the FMO complex, where the system propagator is unitary, we
calculate separately the forward and backward time evolution operators
in MPO form. Using the same compression threshold as in the remaining CCTEMPO steps, we obtain objects with bond dimensions between 8 and
16.
The system propagator MPO describing ideal superradiance is compressed
to bond dimensions 5 to 8.

\section{\label{SM:FMO}
Fenna-Matthews-Olson complex simulations}
\begin{figure*}[h]
\includegraphics[width=\linewidth]{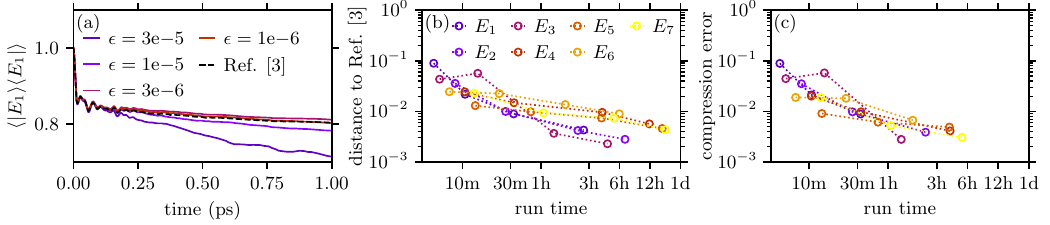}
\caption{\label{fig:FMO_convergence}
Convergence of CCTEMPO for the FMO complex in the main text.
Threshold dependence of the dynamics (a) for the first eigenstate. 
Distance to the reference simulation (b) and compression error with respect to the most converged CCTEMPO data set (c) for the different initial eigenstates as a function of computation time.
}
\end{figure*}
To model the 7-site FMO complex, we use the Hamiltonian 
$H_\text{FMO}$ suggested by Adolphs and Renger~\cite{AdolphsRenger}
\begin{align}
\left(\begin{array}{ccccccc}
200& -87.7& 5.5& -5.9& 6.7& -13.7& -9.9\\
-87.7& 320& 30.8& 8.2& 0.7& 11.8& 4.3\\
5.5& 30.8& 0& -53.5& -2.2& -9.6& 6\\
-5.9& 8.2& -53.5& 110& -70.7& -17.0& -63.3\\
6.7& 0.7& -2.2& -70.7& 270& 81.1& -1.3\\
-13.7& 11.8& -9.6& -17.0& 81.1& 420& 39.7\\
-9.9& 4.3& 6& -63.3& -1.3& 39.7& 230
\end{array}\right)
\end{align}
in units cm$^{-1}$, 
where we shifted all diagonal elements by -12,210 cm$^{-1}$.

The phonon spectral density with 62 peaks per site as well as a broad 
low-frequency background was derived in Ref.~\cite{Raetsep}. 
A succinct summary is also found in the Supplementary Material of Ref.~\cite{coarsegraining}

The convergence of the decay dynamics of the initially prepared first energy eigenstate is shown in
Fig.~\ref{fig:FMO_convergence}(a). With decreasing compression threshold  $\epsilon$ the reference results of Lorenzoni \textit{et al.}~\cite{Lorenzoni2025} are better approximated. The short-time dynamics converges quickly while an accurate description of the long-time dynamics requires small compression thresholds.
To quantitatively assess the convergence, we present in Fig.~\ref{fig:FMO_convergence}(b) the
difference between CCTEMPO calculations for several values of $\epsilon$ and the DAMPF results of Ref.~\cite{Lorenzoni2025} as a function of the computation time. Panel (c) shows the difference with respect to the most converged CCTEMPO result.
We find that CTEMPO converges up to a numerical error of $\sim 10^{-2}$ within $3\,\mathrm{h}-17\,\mathrm{h}$ on an Intel Xeon w3-2435 processor.
It should be noted that the reference result of~\cite{Lorenzoni2025} was obtained using a coarse-graining procedure for the spectral density, which can contribute to the deviations.

\section{\label{SM:superrad}Phonon impact on superradiance of semiconductor quantum dots}

Superradiance occurs when several quantum emitters couple to the common light field with a fixed phase. Coherent coupling can be achieved by placing the emitters at distances smaller than the wavelength of the light. For semiconductor QDs, the most practicable way to achieve coherent coupling is by embedding then into a common wave guide~\cite{superrad_Hallett}. Here, we consider the case of ideal superradiance, where all QDs couple in phase and where the system evolution is governed by the Liouvillian
\begin{align}
\mathcal{L}\hat{\rho}=&\gamma\bigg( L \hat{\rho} L^\dagger 
- \frac 12(L^\dagger L \hat{\rho} + \hat{\rho} L^\dagger L)\bigg),
\end{align}
where $L=\sum_{i=1}^N \sigma^-_i$ is the collective collapse operator and $\sigma^-_i$ denotes the removal of an excitation from QD $i$. 
Starting from an initial state in which all emitters are excited, decay takes place through a ladder of symmetrized bright states. 
The effective rate increases $\sim N$ towards the center of the Dicke ladder, which leads to a characteristic intensity burst as shown in Fig. 4 in the main text. The intensity is evaluated as the negative derivative of the total exciton number in the system 
$I(t)=-\frac{d}{dt}\sum_{i=1}^N \langle \sigma^+_i \sigma^-_i\rangle$.

The QD-phonon interaction is known to be well described by a super-Ohmic spectral density~\cite{Krummheuer2005}
\begin{align}
J(\omega)=\frac{\omega^3}{4\pi^2 \rho_m \hbar c_s^5}\bigg(D_e e^{-\omega^2 a_e^2/(4c_s^2)} - D_h e^{-\omega^2 a_h^2/(4c_s^2)}\bigg)^2,
\end{align}
where we use typical parameters for GaAs-based QDs: electron radius $a_e=4$~nm, hole radius $a_h=a_e/1.15$, mass density $\rho_m=5370$~kg/m$^3$, 
speed of sound $c_s=5110$~m/s, and electron and hole deformation potentials $D_e=7$~eV and $D_h=-3.5$~eV, respectively~\cite{Krummheuer2005}.
For simplicity, we assume that the spectral density as well as the temperature $T=4$~K is the same at all QDs, and that the QDs have identical
transition frequencies. We note that CCTEMPO does not make use of these assumptions and allows one to choose different Hamiltonians and bath spectral densities for each QD. 

Plugging the fully excited initial state, the collective system Liouvillian, and the bath spectral densities into our code, we obtain numerically exact predictions of the time evolution of the emission intensity. We find converged results for a compression threshold $\epsilon=10^{-7}$ while setting the time step to $\Delta t=0.1$~ps for $\gamma^{-1}=500$~ps, and $\Delta t=0.01$~ps for $\gamma^{-1}=50$~ps and  $\gamma^{-1}=5$~ps.

We note that the number $N$ of QDs is limited by working memory of the computer, which is why for these simulations we switch to a workstation with 128~gigabyte of random-access memory. The bonds connecting the backbone to the teeth, and hence the QDs to the phonon environments, retain a small dimension of about $\chi\sim 7$. 
We thus attribute the large overall memory requirements to the fact that the system state is a mixture of strongly entangled states with entanglement distributed over the full network, while the attachment of the non-Markovian environment is only responsible for a moderate increase in the numerical demands.

%

\end{bibunit}
\end{document}